\documentclass{Interspeech}
\interspeechcameraready

\title{FairASR: Fair Audio Contrastive Learning for Automatic Speech Recognition}

\author[affiliation={1}]{Jongsuk}{Kim}
\author[affiliation={1}]{Jaemyung}{Yu}
\author[affiliation={1}]{Minchan}{Kwon}
\author[affiliation={1}]{Junmo}{Kim}

\affiliation{Electrical Engineering}{KAIST}{South Korea}
\email{\{jskpop, jaemyung, kmc0207, junmo.kim\}@kaist.ac.kr}

\usepackage{comment}
\usepackage{booktabs}
\usepackage{array}
\usepackage{subcaption}  
\begin{document}

\maketitle

\begin{abstract}
Large-scale ASR models have achieved remarkable gains in accuracy and robustness. However, fairness issues remain largely unaddressed despite their critical importance in real-world applications. 
In this work, we introduce FairASR, a system that mitigates demographic bias by learning representations that are uninformative about group membership, enabling fair generalization across demographic groups.
Leveraging a multi-demographic dataset, our approach employs a gradient reversal layer to suppress demographic-discriminative features while maintaining the ability to capture generalizable speech patterns through an unsupervised contrastive loss. Experimental results show that FairASR delivers competitive overall ASR performance while significantly reducing performance disparities across different demographic groups.
\end{abstract}

\section{Introduction}

In recent years, large-scale automatic speech recognition (ASR) models, such as Whisper~\cite{radford2023robust}, have achieved remarkable advancements in accuracy and robustness. Despite these advancements, the issue of fairness in ASR systems remains underexplored. 
This is particularly concerning as these systems are increasingly integrated into everyday applications such as voice assistants and call center analytics, where demographic bias can lead to systematic disadvantages for certain user groups.
Recent literature~\cite{trinh2022reducing,feng2021quantifying,liu2022towards} has highlighted performance discrepancies across diverse accents, genders, and sociolects, underscoring the need for more equitable ASR systems.

Several prior studies~\cite{trinh2022reducing,zhang2018mitigating,dheram2022toward} have taken steps toward mitigating biases by proposing model architectures or training strategies that reduce performance gaps between demographic groups. At the same time, the introduction of fairness-focused datasets~\cite{veliche2024towards} has enabled more rigorous evaluation under controlled conditions. These datasets often provide multiple demographics (e.g., accent, gender), allowing for a more fine-grained analysis of how such attributes influence recognition outcomes.

\begin{figure}[t]
    \centering
    \includegraphics[width=\columnwidth]{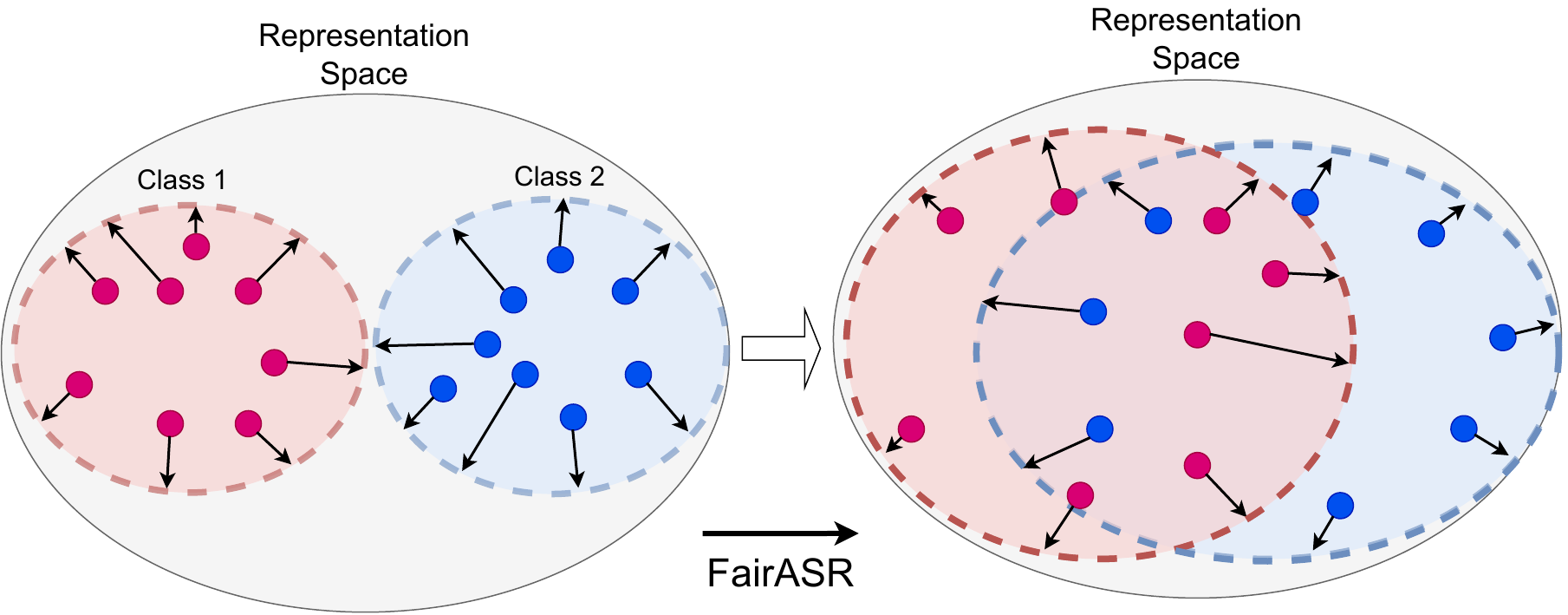}
    \caption{Effect of FairASR on representation space. With standard contrastive learning (left), embeddings cluster by demographic labels. With FairASR (right), such separation is reduced, resulting in fairer representations.}
    \label{fig:intro}
\end{figure}

However, these approaches largely operate beyond the representation learning stage, leaving open the question of how to directly encourage fair and unbiased representations during pretraining.
To address this, we introduce \textbf{FairASR}, Fair Audio Contrastive Learning for Automatic Speech Recognition. 
FairASR leverages multi-demographic supervision during pretraining to enforce demographic-agnostic representations by explicitly reducing demographic separability, as illustrated in Figure~\ref{fig:intro}.
While standard supervised contrastive learning methods~\cite{khosla2020supervised} aim to increase inter-class separability by pulling together representations with the same group, we take a reversed approach that intentionally discourages separation across demographic groups.
We find that this approach preserves ASR accuracy while promoting fairness, offering a pretraining objective more compatible with the nature of speech modeling.

To implement FairASR, we employ a gradient reversal layer~\cite{ganin2015unsupervised} to suppress demographic-discriminative features in the latent space. 
Additionally, we incorporate an unsupervised contrastive loss based on InfoNCE loss~\cite{oord2018representation} to preserve the model’s ability to learn generalizable speech representations. 
Experimental results demonstrate that our method achieves competitive ASR performance while substantially improving fairness, highlighting its potential for equitable speech recognition in real-world applications.

\begin{figure*}[t]
    \centering
    \includegraphics[width=\textwidth]{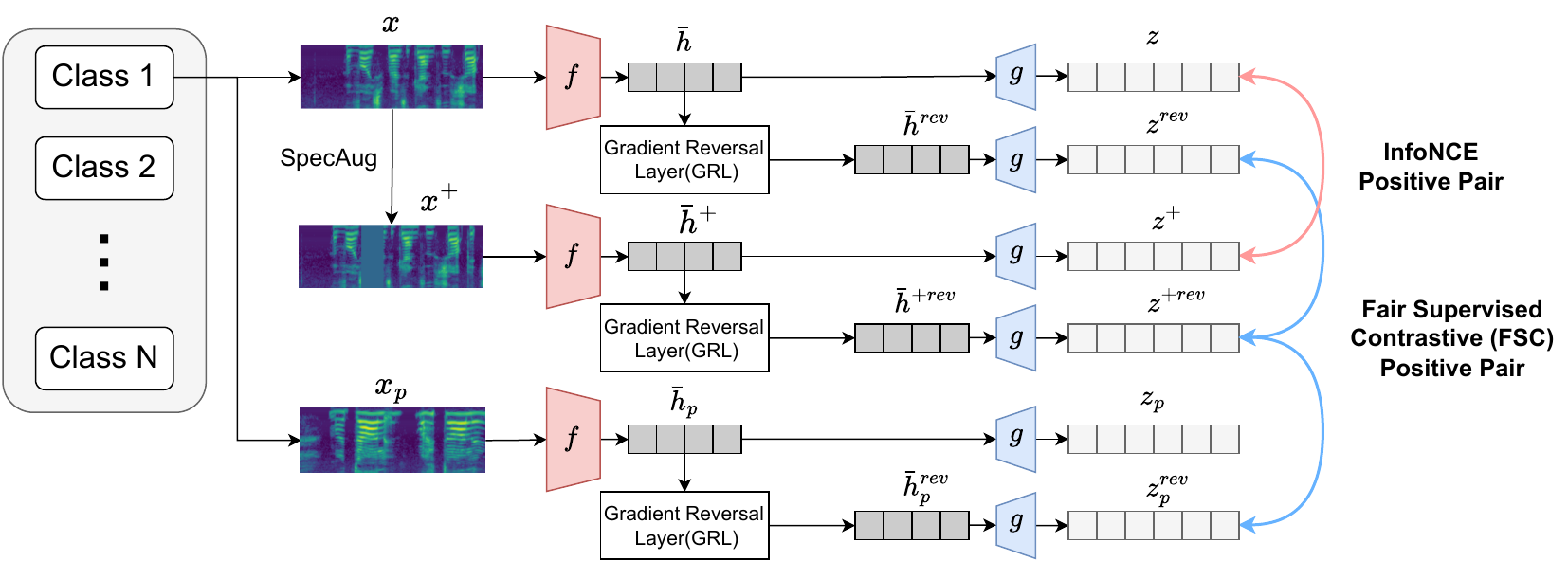}
    \caption{Overview of the proposed adversarial learning framework. SpecAugment is applied to input spectrograms, and a Conformer encoder \( f(\cdot) \) extracts feature embeddings \( \bar{h} \). A gradient reversal layer(GRL) generates adversarial embeddings \( \bar{h}^{\text{rev}} \), which are passed through a shared projection head \( g(\cdot) \). The model is trained with two losses: InfoNCE loss (red) for general representation learning and Fair Supervised Contrastive (FSC) loss (blue) to remove demographic bias. GRL enforces demographic-agnostic representations by reversing gradients, preventing demographic clustering.}
    \label{fig:main}
\end{figure*}

\section{Related Work}

Fairness in automatic speech recognition (ASR) has become a prominent research focus due to concerns about performance biases across different demographics of speakers. Several studies~\cite{tatman2017gender, trinh2022reducing,garnerin2019gender,feng2021quantifying,liu2022towards,sari2021counterfactually,meyer2020artie} have demonstrated that ASR accuracy can vary significantly across demographic attributes such as speaker gender, age, and accent. 
For instance, one study~\cite{tatman2017gender} found that automatic subtitles on YouTube are less accurate for female speakers and particular dialects, while another study~\cite{koenecke2020racial} highlighted significant racial disparities in commercial ASR systems.
The recently introduced Fair-Speech dataset~\cite{veliche2024towards} was explicitly designed to support fairness assessments by encompassing a wide range of demographics.
This is used as a key metric for assessing bias and highlights the vulnerabilities of existing ASR models.

To mitigate demographic biases in ASR models, recent studies have proposed a variety of training methods.
One approach~\cite{zhang2018mitigating} utilizes adversarial learning, in which an adversary attempts to predict the demographic information of the speaker from the latent features of the ASR model, encouraging the development of invariant representations. 
Another method~\cite{trinh2022reducing} involves domain-adaptive fine-tuning, where a pre-trained ASR model is refined using speech data from underrepresented groups. 
In \cite{dheram2022toward}, unsupervised speaker embeddings combined with oversampling and cohort membership modeling effectively reduce ASR performance disparities.

\section{Method}
In this work, we propose a demographic-invariant representations learning method by combining self-supervised contrastive learning and adversarial supervised contrastive learning. Our method explicitly utilizes demographic labels during training, ensuring that learned representations remain robust, discriminative, and fair across diverse populations. The overall framework consists of the following steps.

\subsection{Pre-processing}
The audio waveform is converted into a Mel-spectrogram using 80 Mel filter banks at a sample rate of 16 kHz. 
We generate an augmented version of each original spectrogram in a batch using SpecAugment~\cite{park2019specaugment}. 
For a batch of $N$, original waveforms are expanded to a set $\{x_i\}_{i=1}^{2N}$ containing both the original and augmented samples.
Each sample $x_i$ is associated with a demographic group label $d_i$ (e.g., age, accent, or regional background) and transcription $y_i$. The augmented samples inherit the same demographic group label as their original counterparts.

\subsection{Architecture}
\subsubsection{Backbone Encoder}
We use a Conformer encoder~\cite{gulati2020conformer} as the backbone feature extractor. Given an input Mel-spectrogram $x_i$, the encoder generates a sequence of features: $h=f(x)\in \mathbb{R}^{2N\times D \times T}$, where $D$ is the feature dimension per frame and $T$ is the number of time frames. 
Then, we apply mean pooling over time to obtain a single vector for each sample: $\bar{h}=\operatorname{MeanPool}(h)$.
This results in a batch of pooled feature vectors $\{ \bar{h}_i \}_{i=1}^{2N}$, where $\bar{h}_i$ representing either an original or an augmented waveform.

\subsubsection{Embedding with Gradient Reversal Layer}
To encourage demographic-agnostic features, we employ gradient reversal layer (GRL) to adversarially prevent demographic clustering in the learned representations. We obtain a $\bar{h}^{\text{rev}}$ by passing it through the GRL: $\bar{h}^{\text{rev}} =\operatorname{GRL}(\bar{h})$.
The GRL does not alter the values of $\bar{h}$ but reverses the direction of their gradients during backpropagation, which helps to learn features that are insensitive to demographic labels. Both $\bar{h}$ and $\bar{h}^{\text{rev}}$ are then fed into the same projection head $g(\cdot)$, which is implemented as a two-layer MLP with a non-linear activation function:
\begin{equation}
    z=g(\bar{h}), \quad z^{\text{rev}} = g(\bar{h}^{\text{rev}}),
\end{equation}
where $z, z^{rev} \in \mathbb{R}^{2N\times D'}$ and $D'$ is the embedding dimension.

\subsection{Loss Functions}
\subsubsection{InfoNCE Loss for Representation Learning}
To maintain the high quality of the feature, we apply the InfoNCE loss~\cite{oord2018representation} on the embeddings $z$.
For each original sample $i$ (with embedding $z_i$), the positive counterpart is its augmented version $i^+$ (with embedding $z_i^+$), and all other $2N-2$ samples in the batch serve as negatives. The loss is defined as:
\begin{equation}\label{eq:info}
    \mathcal{L}_{\text{InfoNCE}} = - \sum_{i=1}^{2N} \log \frac{\exp\left(z_i \cdot z_i^+ / \tau\right)}{\sum_{j\neq i} \exp\left(z_i \cdot z_j / \tau\right)},
\end{equation}
where $\cdot$ symbol denotes the inner product and $\tau$ is temperature.

\subsubsection{Fair Supervised Contrastive Loss for Bias Removal}
To explicitly prevent demographic clustering, we employ a supervised contrast loss function~\cite{khosla2020supervised} to the embeddings passed through the gradient reversal layer, $z^{\text{rev}}$.
For an anchor sample $i$ with a demographic group label $d_i$, we consider all other samples $p$ in the batch that share the same demographic group as positive examples. Let $P(i)$ be the set of indices of these same-group samples (including the augmented version of $i$ itself). The fair supervised contrastive (FSC) loss for anchor $i$ is given by:
\begin{align}
    &\mathcal{L}_{\text{FSC}}=-\sum_{i=1}^{2N} \frac{1}{|P(i)|} \sum_{p \in P(i)} \log \frac{\exp \left(z_i^{\text{rev}} \cdot z_p^{\text{rev}}/ \tau\right)}{\displaystyle\sum_{a \neq i} \exp \left(z_i^{\text{rev}} \cdot z_a^{\text{rev}}/ \tau\right)}, \\
    &\text{where} \quad P(i) = \{j | d_j = d_i \text{~and~} j \neq i \}.
\end{align}
This loss pulls together the embeddings $z_i^{\text{rev}}$ with others $z_p^{\text{rev}}$ from the same demographic label while pushing apart embeddings $z_a^{\text{rev}}$. On the other hand, through the GRL and SupCon pathways, the encoder is encouraged to produce embeddings that do not carry demographic-discriminative information.

\subsubsection{Overall Objective} 
The overall loss is computed by aggregating both the InfoNCE and FSC losses over all samples in the batch:
\begin{equation}\label{eq:fairasr}
    \mathcal{L}_{\text {FairASR}}= \mathcal{L}_{\text{InfoNCE}} + \lambda \mathcal{L}_{\text{FSC}},
\end{equation}
where $\lambda$ is a hyperparameter that controls the balance between maintaining representation quality and enforcing demographic invariance. By minimizing this FairASR loss, the Conformer encoder $f$ is trained to produce rich audio representations that are informative and demographic-agnostic.

\subsection{ASR Fine-tuning}
After pre-training with FairASR, the encoder learns representations that are invariant to demographic attributes.  
During fine-tuning, we utilize the task label \( y_i \) for Automatic Speech Recognition (ASR). Given an input \( x \), the encoder extracts a sequence of frame-level representations \( h = (h_1, \dots, h_T) \), which is passed through a single-layer LSTM~\cite{graves2012long} for temporal modeling, followed by a linear projection to the vocabulary space to generate transcriptions.

The ASR model is then trained using Connectionist Temporal Classification (CTC) loss~\cite{graves2006connectionist}:  
\begin{equation}
\mathcal{L}_{\text{CTC}} = -\log P(y \mid x),
\end{equation}
where \( P(y\mid x) \) is the total probability summed over all valid CTC alignments between the input and output sequences.  
Since fine-tuning is performed on FairASR-pretrained representations, the model preserves demographic invariance, which contributes to fairer transcription performance across demographic groups.

\begin{table}[!t]
\large
 \caption{Data distribution and performances with varying loss function, compared to the commercial baseline. Ethnic groups are abbreviated for clarity: Asn. (South Asian or Asian American), Blk. (Black or African American), Hsp. (Hispanic, Latino, or Spanish), Mea. (Middle Eastern or North African), Nai. (Native American, American Indian, or Alaska Native), Nhp. (Native Hawaiian or Other Pacific Islander), Wht. (White).}
  \centering
  \resizebox{\columnwidth}{!}{%
  \begin{tabular}{l|c|c|c|c|c}
    \toprule
    & \multicolumn{2}{c|}{\# data} & \multicolumn{2}{c|}{Loss function} & Commercial\\
    \cmidrule(r){2-6}
    & train & test & InfoNCE & FairASR & Whisper\\
    \midrule
    \midrule
    \multicolumn{2}{l}{\emph{Age}} \\
    \midrule
    18 - 22                                              & 3398  & 362   & 5.24 & 5.54 & 4.46 \\
    23 - 30                                              & 3616  & 396   & 6.97 & 6.93 & 4.62\\
    31 - 45                                              & 11222 & 1255  & 7.54 & 8.08 & 7.48\\
    46 - 65                                              & 4710  & 537   & 5.10 & 5.59 & 3.65\\
    \midrule
    WER gap (\%)                                    & -     & -     & 32.4 & \textbf{31.4} & 51.2\\
    \midrule
    \midrule
    \multicolumn{2}{l}{\emph{Gender}} \\
    \midrule
    Female                                               & 12386 & 1368  & 5.31 & 5.68 & 3.86\\
    Male                                                 & 10560 & 1182  & 8.00 & 8.47 & 7.91\\
    \midrule
    WER gap (\%)                                    & -     & -     & 33.6 & \textbf{32.9} & 51.2\\
    \midrule
    \midrule
    \multicolumn{2}{l}{\emph{Ethnicity}} \\
    \midrule
    Asn.                                              & 3389  & 383 & 5.41 & 5.56 & 3.70\\
    Blk.                                              & 6787  & 784 & 8.87 & 9.52 & 9.52\\
    Hsp.                                           & 2407  & 260 & 4.14 & 4.86 & 3.90\\
    Mea.                                     & 646   & 76  & 6.84 & 6.47 & 5.08\\
    Nai.                                    & 3963  & 407 & 6.66 & 6.74 & 4.12\\
    Nhp.                                    & 861   & 101 & 7.13 & 7.33 & 3.84\\
    Wht.                                               & 4893  & 539 & 5.07 & 5.44 & 3.96\\
    \midrule
    WER gap (\%)                                    & -     & -   & 53.3 & \textbf{48.9} & 61.1\\
    \midrule
    \midrule
    \multicolumn{2}{l}{\emph{Socioeconomic}} \\
    \midrule
    Low                                 & 12751 & 1419 & 6.33 & 6.64 & 5.40\\
    Medium                           & 8566  & 950  & 7.20 & 7.58 & 6.38\\
    Affluent                             & 1629  & 181  & 5.78 & 6.41 & 3.62\\
    \midrule
    WER gap (\%)                                    & -     & -    & 19.7 & \textbf{15.4} & 43.3\\
    \midrule
    \midrule
    \multicolumn{2}{l}{\emph{First language}} \\
    \midrule
    English                                              & 18729 & 2080 & 6.82 & 7.19 & 6.11\\
    Non-English                                          & 4217  & 470  & 5.68 & 6.31 & 3.78\\
    \midrule
    WER gap (\%)                                    & -     & -    & 16.7 & \textbf{12.2} & 46.8\\
    \bottomrule
  \end{tabular}
  }
  \label{tab:data}
\end{table}

\section{Experimental Results}
\begin{figure}[t]
    \centering
    \includegraphics[width=\columnwidth]{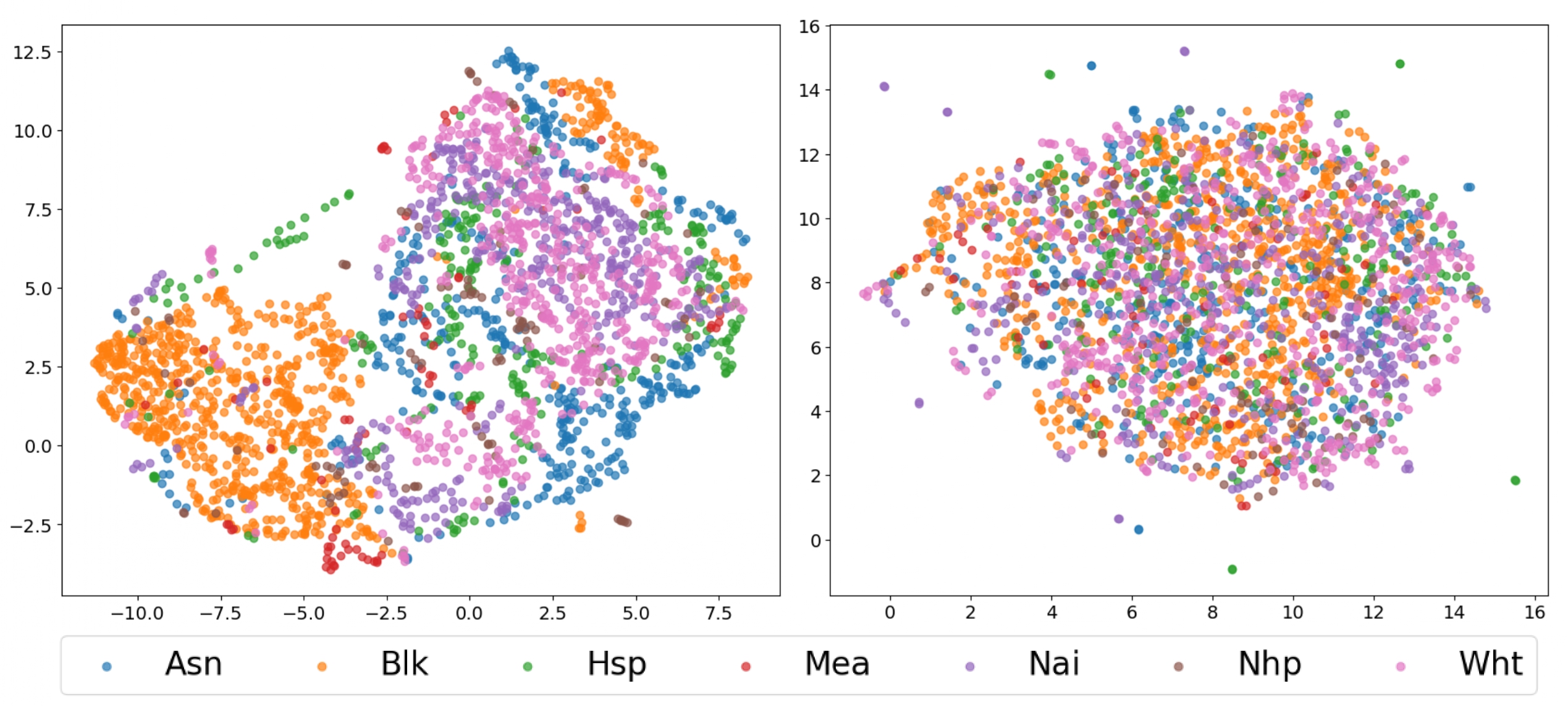}
    \caption{UMAP visualization of representations. Left: InfoNCE-only training shows demographic clustering. Right: FairASR promotes more mixed and fair representations.}
    \label{fig:tsne}
\end{figure}

\subsection{Experimental Settings}
\subsubsection{Dataset}
We use the FairSpeech dataset~\cite{veliche2024towards}, the most recently published dataset addressing fairness in ASR. 
Since the dataset contains some excessively long audio segments, we restrict the maximum length to 280k samples at a 16kHz sample rate (17.5 seconds) to ensure stable training. 
This filtering reduces the number of samples from 26,472 to 22,946. 
The dataset is partitioned into training and test sets, maintaining a consistent label distribution.
The training set is used for pre-training and ASR fine-tuning, while the test set is reserved exclusively for evaluation. 
The detailed distribution of training and test samples per label is provided in Table~\ref{tab:data}.

\subsubsection{Evaluation Metrics}
We evaluate model performance based on the Word Error Rate (WER), which is calculated as follows:
$\text{WER} = (S + D + I)/N_T,$ where $S, D$, and $I$ represent the number of substitutions, deletions, and insertions, respectively, and $N_T$ is the total number of words in the reference transcription.
To assess fairness, we compute the WER gap across different speaker cohorts. The WER gap is defined as: $\text{WER gap} = (\text{WER}_{\text{max}} - \text{WER}_{\text{min}})/\text{WER}_{\text{max}}$,
where $\text{WER}_{\text{max}}$ and $\text{WER}_{\text{min}}$ denote the highest and lowest WER observed among the cohorts, respectively. A lower WER gap indicates more balanced performance across groups, suggesting a fairer ASR system.

\subsubsection{Implementation Details}
We employ the Conformer small model~\cite {gulati2020conformer} as our backbone, initializing it with weights obtained from supervised training on LibriSpeech. During pre-training, we use the AdamW optimizer with a learning rate of $1\times 10^{-4}, (\beta_1,\beta_2)=(0.9,0.999)$, and a weight decay of $0.01$.
Training is conducted for 100 epochs using a cosine annealing scheduler. The temperature parameter is set to 0.2, and we experiment with $\lambda$ values of 0.1. For data augmentation, we utilize SpecAugment with both time masking and frequency masking.
Both pre-training and ASR fine-tuning are performed with a total batch size of 64. The ASR fine-tuning settings follow the standard Conformer training configuration, except that no data augmentation is applied.

\subsection{Main Results}
As reported in prior work~\cite{veliche2024towards}, the Whisper large-v2 model achieves strong overall WER performance due to large-scale training on diverse data. 
While this model achieves strong overall WER performance due to extensive data and training, its high WER gap across different speaker cohorts reveals significant shortcomings in fairness.
We further compare models trained solely with the InfoNCE loss (Eq.~\ref{eq:info}) to those trained with the FairASR loss (Eq.~\ref{eq:fairasr}).
As shown in Table~\ref{tab:data}, the FairASR-trained model significantly reduces the WER gap across demographic groups, despite exhibiting a slightly higher WER compared to the InfoNCE-only model.
These results demonstrate that FairASR effectively enhances fairness by mitigating performance disparities across demographic groups.

Furthermore, we analyze the representations using UMAP, as shown in Figure~\ref{fig:tsne}.
The InfoNCE-trained model yields features that form distinct clusters by demographic group, indicating that group-specific information is preserved.
In contrast, FairASR produces more overlapping distributions, suggesting demographic invariance in the learned space.


\begin{table}[!t]
 \caption{Comparison of embedding space sharing and balance parameter variation in overall performance. ``Ind." denotes independent.}
  \centering
  \scriptsize
  \renewcommand{\arraystretch}{0.8}
  \resizebox{\columnwidth}{!}{%
  \begin{tabular}{l|c|c|c|c}
    \toprule
    Embedding space & Ind. & Share & Share & Share \\
    \midrule
    Balance parameter($\lambda$)         & 0.1  & 0.1  & 0.01 &  1   \\
    \midrule
    \midrule
    \multicolumn{1}{l}{\emph{Age}} \\
    \midrule
    18 - 22 & 5.29 & 5.54 & 5.76 & 5.29 \\
    23 - 30 & 6.95 & 6.93 & 7.46 & 7.19 \\
    31 - 45 & 8.35 & 8.08 & 8.10 & 7.85 \\
    46 - 65 & 5.51 & 5.59 & 5.49 & 5.16 \\
    \midrule
    WER gap (\%)                                   & 36.6 & \textbf{31.4} & 32.2 & 34.3 \\
    \midrule
    \midrule
    \multicolumn{1}{l}{\emph{Gender}} \\
    \midrule
    Female  & 5.65 & 5.68 & 5.64 & 5.46 \\
    Male    & 8.68 & 8.47 & 8.71 & 8.26 \\
    \midrule
    WER gap (\%)                                   & 34.9 & \textbf{32.9} & 35.2 & 33.9 \\
    \midrule
    \midrule
    \multicolumn{1}{l}{\emph{Ethnicity}} \\
    \midrule
    Asn.   & 5.66 & 5.56 & 5.33 & 5.43 \\
    Blk.   & 9.35 & 9.52 & 9.73 & 9.22 \\
    Hsp.& 4.66 & 4.86 & 5.25 & 5.06 \\
    Mea.  & 7.46 & 6.47 & 6.47 & 7.21 \\
    Nai. & 6.77 & 6.74 & 6.90 & 6.34 \\
    Nhp. & 7.13 & 7.33 & 7.93 & 7.43 \\
    Wht.   & 5.49 & 5.44 & 5.40 & 5.11 \\
    \midrule
    WER gap (\%)                                    & 50.2 & 48.9 & 46.0   & \textbf{45.1} \\
    \midrule
    \midrule
    \multicolumn{1}{l}{\emph{Socioeconomic background}} \\
    \midrule
    Low      & 6.69 & 6.64 & 6.78 & 6.45 \\
    Medium   & 7.88 & 7.58 & 7.75 & 7.43 \\
    Affluent & 6.30 & 6.41 & 6.35 & 6.30\\
    \midrule
    WER gap (\%)                                    & 20.1 & 15.4 & 18.1 & \textbf{15.2} \\
    \midrule
    \midrule
    \multicolumn{1}{l}{\emph{First language}} \\
    \midrule
    English    & 7.24 & 7.19 & 7.37 & 6.99 \\
    Non-English& 6.28 & 6.31 & 6.00 & 6.20 \\
    \midrule
    WER gap (\%)                                    & 13.3 & 12.2 & 18.6 & \textbf{11.3} \\
    \midrule
    \midrule
    \textbf{Total WER}               & 7.12 & 6.97 &7.11 &\textbf{6.88}\\
    \bottomrule
  \end{tabular}
  }
  \label{tab:embedding}
\end{table}

\subsection{Ablation Studies}
\subsubsection{Embedding Space}
We investigate whether computing the InfoNCE loss and the FSC loss in a shared or independent embedding space is more advantageous. This experiment examines whether jointly optimizing feature quality and demographic-agnostic representations in the same space leads to better performance. For the independent configuration, we implemented separate projection heads for $z$ and $z^{\text{rev}}$. As shown in Table~\ref{tab:embedding} (columns 1 and 2), while maintaining the same $\lambda$ values, the shared embedding space configuration results in a significantly lower WER gap compared to the independent configuration. Additionally, the overall WER is lower when using a shared space. These results demonstrate that sharing the embedding space for both loss functions is beneficial, enhancing both the representation quality and the fairness of the ASR system.

\subsubsection{Balance Parameter}
Table~\ref{tab:embedding} (columns 2–4) shows how performance varies with changes in \(\lambda\). When \(\lambda\) is set too low (e.g., \(\lambda = 0.01\)), the model exhibits suboptimal performance in both the WER and WER gap. In contrast, when \(\lambda\) is within an appropriate range, we observe a trade-off between representation quality and fairness, reflecting the balance between these aspects depending on the label distribution.

\section{Conclusion}
In this work, we introduce FairASR, a fair contrastive learning framework that mitigates demographic bias at the representation learning stage of ASR models.  
Unlike most prior work that addresses fairness post hoc, FairASR directly encourages demographic invariance during pretraining.  
Extensive experiments show that FairASR consistently reduces WER gaps across diverse demographic categories, supported by both quantitative metrics and qualitative analyses.  
We hope that this work inspires further research on fairness-aware representation learning and contributes to the development of more equitable and inclusive ASR systems.

\section{Acknowledgements}
This work was supported by Institute of Information \& communications Technology Planning \& Evaluation (IITP) grant funded by the Korea government(MSIT) (No.RS-2022-II220184, Development and Study of AI Technologies to Inexpensively Conform to Evolving Policy on Ethics)
\bibliographystyle{IEEEtran}
\bibliography{mybib}

\begin{thebibliography}{10}
\providecommand{\url}[1]{#1}
\csname url@samestyle\endcsname
\providecommand{\newblock}{\relax}
\providecommand{\bibinfo}[2]{#2}
\providecommand{\BIBentrySTDinterwordspacing}{\spaceskip=0pt\relax}
\providecommand{\BIBentryALTinterwordstretchfactor}{4}
\providecommand{\BIBentryALTinterwordspacing}{\spaceskip=\fontdimen2\font plus
\BIBentryALTinterwordstretchfactor\fontdimen3\font minus \fontdimen4\font\relax}
\providecommand{\BIBforeignlanguage}[2]{{%
\expandafter\ifx\csname l@#1\endcsname\relax
\typeout{** WARNING: IEEEtran.bst: No hyphenation pattern has been}%
\typeout{** loaded for the language `#1'. Using the pattern for}%
\typeout{** the default language instead.}%
\else
\language=\csname l@#1\endcsname
\fi
#2}}
\providecommand{\BIBdecl}{\relax}
\BIBdecl

\bibitem{radford2023robust}
A.~Radford, J.~W. Kim, T.~Xu, G.~Brockman, C.~McLeavey, and I.~Sutskever, ``Robust speech recognition via large-scale weak supervision,'' in \emph{International conference on machine learning}.\hskip 1em plus 0.5em minus 0.4em\relax PMLR, 2023, pp. 28\,492--28\,518.

\bibitem{trinh2022reducing}
V.~A. Trinh, P.~Ghahremani, B.~King, J.~Droppo, A.~Stolcke, and R.~Maas, ``Reducing geographic disparities in automatic speech recognition via elastic weight consolidation,'' \emph{arXiv preprint arXiv:2207.07850}, 2022.

\bibitem{feng2021quantifying}
S.~Feng, O.~Kudina, B.~M. Halpern, and O.~Scharenborg, ``Quantifying bias in automatic speech recognition,'' \emph{arXiv preprint arXiv:2103.15122}, 2021.

\bibitem{liu2022towards}
C.~Liu, M.~Picheny, L.~Sar{\i}, P.~Chitkara, A.~Xiao, X.~Zhang, M.~Chou, A.~Alvarado, C.~Hazirbas, and Y.~Saraf, ``Towards measuring fairness in speech recognition: Casual conversations dataset transcriptions,'' in \emph{ICASSP 2022-2022 IEEE International Conference on Acoustics, Speech and Signal Processing (ICASSP)}.\hskip 1em plus 0.5em minus 0.4em\relax IEEE, 2022, pp. 6162--6166.

\bibitem{zhang2018mitigating}
B.~H. Zhang, B.~Lemoine, and M.~Mitchell, ``Mitigating unwanted biases with adversarial learning,'' in \emph{Proceedings of the 2018 AAAI/ACM Conference on AI, Ethics, and Society}, 2018, pp. 335--340.

\bibitem{dheram2022toward}
P.~Dheram, M.~Ramakrishnan, A.~Raju, I.-F. Chen, B.~King, K.~Powell, M.~Saboowala, K.~Shetty, and A.~Stolcke, ``Toward fairness in speech recognition: Discovery and mitigation of performance disparities,'' in \emph{Proc. Interspeech 2022}, 2022, pp. 1268--1272.

\bibitem{veliche2024towards}
I.-E. Veliche, Z.~Huang, V.~Ayyat~Kochaniyan, F.~Peng, O.~Kalinli, and M.~L. Seltzer, ``Towards measuring fairness in speech recognition: Fair-speech dataset,'' in \emph{Proc. Interspeech 2024}, 2024, pp. 1385--1389.

\bibitem{khosla2020supervised}
P.~Khosla, P.~Teterwak, C.~Wang, A.~Sarna, Y.~Tian, P.~Isola, A.~Maschinot, C.~Liu, and D.~Krishnan, ``Supervised contrastive learning,'' \emph{Advances in neural information processing systems}, vol.~33, pp. 18\,661--18\,673, 2020.

\bibitem{ganin2015unsupervised}
Y.~Ganin and V.~Lempitsky, ``Unsupervised domain adaptation by backpropagation,'' 2015.

\bibitem{oord2018representation}
A.~v.~d. Oord, Y.~Li, and O.~Vinyals, ``Representation learning with contrastive predictive coding,'' \emph{arXiv preprint arXiv:1807.03748}, 2018.

\bibitem{tatman2017gender}
R.~Tatman, ``Gender and dialect bias in youtube’s automatic captions,'' in \emph{Proceedings of the first ACL workshop on ethics in natural language processing}, 2017, pp. 53--59.

\bibitem{garnerin2019gender}
M.~Garnerin, S.~Rossato, and L.~Besacier, ``Gender representation in french broadcast corpora and its impact on asr performance,'' in \emph{Proceedings of the 1st international workshop on AI for smart TV content production, access and delivery}, 2019, pp. 3--9.

\bibitem{sari2021counterfactually}
L.~Sar{\i}, M.~Hasegawa-Johnson, and C.~D. Yoo, ``Counterfactually fair automatic speech recognition,'' \emph{IEEE/ACM Transactions on Audio, Speech, and Language Processing}, vol.~29, pp. 3515--3525, 2021.

\bibitem{meyer2020artie}
J.~Meyer, L.~Rauchenstein, J.~D. Eisenberg, and N.~Howell, ``Artie bias corpus: An open dataset for detecting demographic bias in speech applications,'' in \emph{Proceedings of the twelfth language resources and evaluation conference}, 2020, pp. 6462--6468.

\bibitem{koenecke2020racial}
A.~Koenecke, A.~Nam, E.~Lake, J.~Nudell, M.~Quartey, Z.~Mengesha, C.~Toups, J.~R. Rickford, D.~Jurafsky, and S.~Goel, ``Racial disparities in automated speech recognition,'' \emph{Proceedings of the national academy of sciences}, vol. 117, no.~14, pp. 7684--7689, 2020.

\bibitem{park2019specaugment}
D.~S. Park, W.~Chan, Y.~Zhang, C.-C. Chiu, B.~Zoph, E.~D. Cubuk, and Q.~V. Le, ``Specaugment: A simple data augmentation method for automatic speech recognition,'' \emph{arXiv preprint arXiv:1904.08779}, 2019.

\bibitem{gulati2020conformer}
A.~Gulati, J.~Qin, C.-C. Chiu, N.~Parmar, Y.~Zhang, J.~Yu, W.~Han, S.~Wang, Z.~Zhang, Y.~Wu \emph{et~al.}, ``Conformer: Convolution-augmented transformer for speech recognition,'' \emph{arXiv preprint arXiv:2005.08100}, 2020.

\bibitem{graves2012long}
A.~Graves and A.~Graves, ``Long short-term memory,'' \emph{Supervised sequence labelling with recurrent neural networks}, pp. 37--45, 2012.

\bibitem{graves2006connectionist}
A.~Graves, S.~Fernández, F.~Gomez, and J.~Schmidhuber, ``Connectionist temporal classification: Labelling unsegmented sequence data with recurrent neural networks,'' in \emph{Proceedings of the 23rd International Conference on Machine Learning}, 2006, pp. 369--376.

\end{thebibliography}

\end{document}